%% file: skeleton.tex
\documentclass[a4paper,11pt]{article}
\usepackage{pos}
\usepackage{multirow}
\usepackage{lineno}

\input{extradef}

\title{CKM and $CPV$ measurements in the beauty and charm sector}

\author*[a]{Lei Hao}

\onbehalf{on behalf of the \lhcb collaboration}

\affiliation[a]{University of Chinese Academy of Sciences,\\
  Yuquan Rd, Beijing, China}

\emailAdd{hao.lei@cern.ch}

\abstract{
Measurements of CKM elements and \CP-violating observables are sensitive to revealing effects beyond the SM. These proceedings discuss four recent \lhcb results. The first one presents \CP violation measurements in $B^0_s$ decays. Tree-level measurements of the CKM angle $\gamma$ are one of the most important tests of $\CP$ violation in the SM. Additionally, the second one presents the results of a recent analysis of $B^{\pm} \rightarrow [h'^+h'^-\pi^+\pi^-]_Dh^{\pm}$ ($h=K,\pi$).
The third one discusses the combination of previous \lhcb $\gamma$ measurements except for the presented results. Achieved precision of the \lhcb result: $(67 \pm 4)^{\circ}$ dominates the world average.
Charm physics serves as a unique probe to test the flavor sector in the SM. The fourth one shows the search for $\CP$ violation in the multi-body $D$ decays.}

\FullConference{The Eleventh Annual Conference on Large Hadron Collider Physics (LHCP2023)\\
 22-26 May 2023\\
 Belgrade, Serbia\\}

\begin{document}
\maketitle

\section{Introduction} 
The CKM matrix elements (denoted as $V_{ij}$) represent the strength of flavour-changing weak interactions~\cite{PhysRevLett.10.531}. 
Charge parity (\CP) symmetry violation (\CP violation) involves the phases of CKM elements~\cite{Rosner:1996yb,ParticleDataGroup:2018ovx}. Many measurements of \CP-violating observables serve to improve the determination of the CKM elements or to reveal effects beyond the Standard Model (SM). The unitarity of the CKM matrix can be checked. Deviation from the CKM unitarity can be a signal of new physics (NP) beyond the SM.

\section{\texorpdfstring{$\CP$}{CP} violation in \texorpdfstring{$B^0_s$}{Bs0} decays}

$\CP$ violation can originate from the interference of the decay amplitude and 
$\phi_s = -2 \text{arg}[-(V_{ts}V_{tb}^*) \/ (V_{cs}V_{cb}^*) ]$ can be predicted very precisely within the SM~\cite{Charles:2020dfl, UTfit:2022hsi} assuming no NP contributions, so any deviation would be a strong hint of some effects beyond the SM (BSM). Another physical quantity is the width difference. The width difference is predicted less precisely~\cite{Artuso:2015swg}. NP contribution also influences the width difference.

\begin{figure}
    \centering
    \includegraphics[width=0.45\textwidth]{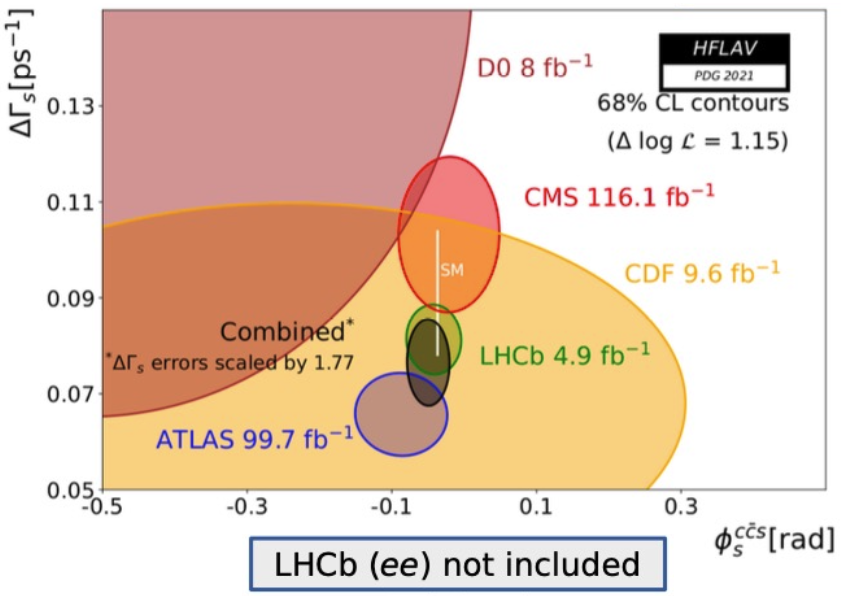}
    \includegraphics[width=0.45\textwidth]{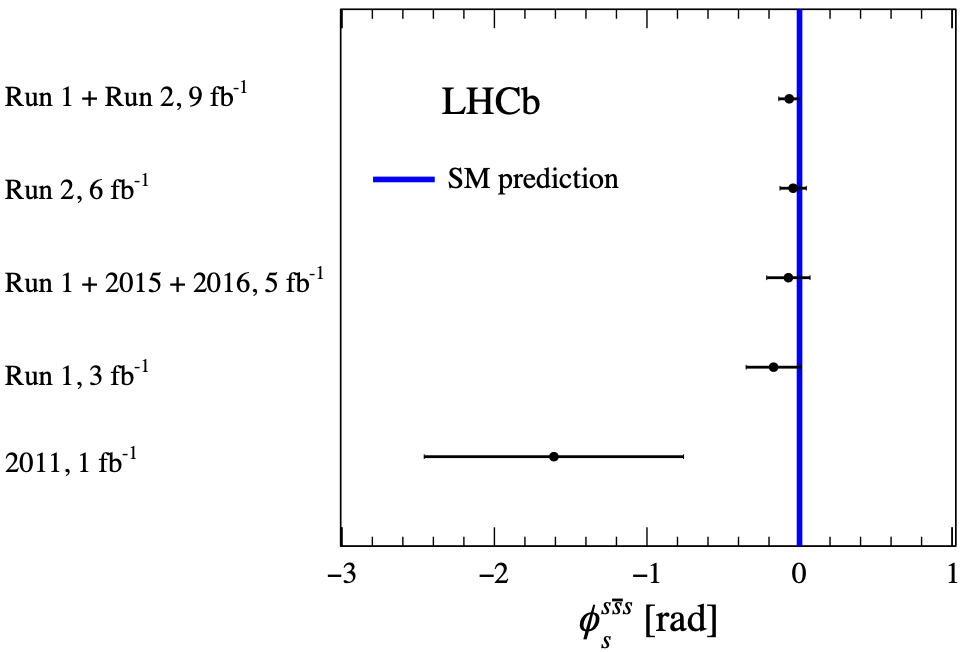}
    \caption{Confidence level contours~\cite{HFLAV:2022esi} (left) of \atlas, \cms, \cdf, \dzero and \lhcb measurements~\cite{CMS:2020efq, ATLAS:2020lbz, LHCb:2019nin, LHCb:2021wte}, combined contour in a black solid line and shaded area, as well as the SM predictions in a very thin white rectangle. Comparison~\cite{LHCb:2023exl} (right) of $\phi_s^{s\bar{s}s}$ measurements from various analyses~\cite{LHCb:2013xyz, LHCb:2014vmq, LHCb:2019jgw} by the \lhcb collaboration. The vertical band indicates the SM prediction~\cite{Cheng:2009mu, Beneke:2006hg, Raidal:2002ph}. 
    }
    \label{fig:phismeasurement}
\end{figure}

There are many measurements of $\phi_s$ from \lhcb, \atlas and \cms ~\cite{ATLAS:2020lbz, CMS:2020efq, LHCb:2021wte, LHCb:2019nin}. In \lhcb, both same-side and opposite-side taggers are used, while in \cms and \atlas, only the opposite-side tagger is utilized. The helicity basis for the angular definition is employed in \lhcb, whereas the transversity basis for the angular definition is used for \atlas and \cms. The flavor-tagged time-dependent angular analysis is performed for the signal extracted from the data by all three collaborations. The measurements from \atlas ~\cite{ATLAS:2020lbz}, \cms ~\cite{CMS:2020efq}, and \lhcb ~\cite{LHCb:2021wte, LHCb:2019nin} are consistent with each other. 
The combined result is consistent with SM predictions, as shown in the left plot of Fig.~\ref{fig:phismeasurement}.

Flavor-changing neutral current decays of $B$ mesons are highly sensitive to NP. 
The $B^0_s\rightarrow \phi\phi$ decay, which proceeds via a $b\rightarrow s\bar{s}s$ transition, is a benchmark channel. 
Time-dependent \CP violation arises from the interference between the direct decay and the decay after $B^0_s$ and $\bar{B}_s^0$ mixing, which can be characterized by the phase $\phi_s^{s\bar{s}s}$ and the parameter $|\lambda|$, which is related to direct \CP violation~\cite{LHCb:2023exl}. 
NP contributions could significantly alter the values predicted by the SM. 
In addition, the final state has three polarization states, and NP may depend on the $B_s^0$ polarisation. 
The flavor-tagged time-dependent angular analysis is performed to measure $\phi_s^{s\bar{s}s}= -0.042 \pm 0.075$ and $|\lambda|=1.004 \pm 0.030$, which is consistent with and supersedes the previous measurement~\cite{LHCb:2019jgw}, and agrees with the SM expectation, as shown in the right plot of Fig.~\ref{fig:phismeasurement}.

\section{Direct measurement of the CKM angle \texorpdfstring{$\gamma$}{gamma}}

\begin{figure}
    \centering
    \includegraphics[width=0.7\textwidth]{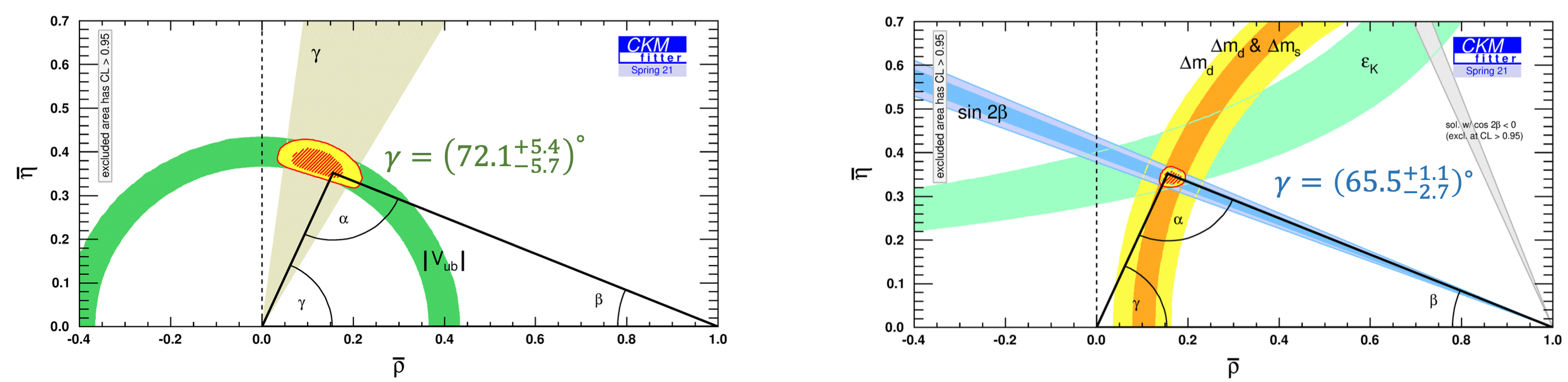}
    \caption{Comparison between direct measurements (left) of the CKM angle $\gamma$~\cite{Brod:2013sga} and indirect measurements (right) of the CKM angle $\gamma$~\cite{Charles:2020dfl}.}
    \label{fig:CKMcomp}
\end{figure}
Direct measurements of the angle $\gamma \equiv \text{arg}[-V_{ud}V_{ub}^*\/V_{cd}V_{cb}^*]$ can be accessible at tree level and serve as benchmarks of the SM. Assuming no NP at tree level processes, the theoretical uncertainties are negligible in such direct measurements~\cite{Brod:2013sga}. In indirect measurements, there are some inputs including loop processes, and the angle $\gamma$ is obtained from the global fit to the unitary triangle, assuming a closed triangle. The comparison between direct measurements and indirect measurements is shown in the left and right plots in Fig.~\ref{fig:CKMcomp}~\cite{Charles:2020dfl}. Loop processes are expected to be sensitive to a possible BSM signature. A discrepancy between direct and indirect measurements would be a clear sign of NP. 

\begin{figure}
    \centering
    \includegraphics[width=0.33\textwidth]{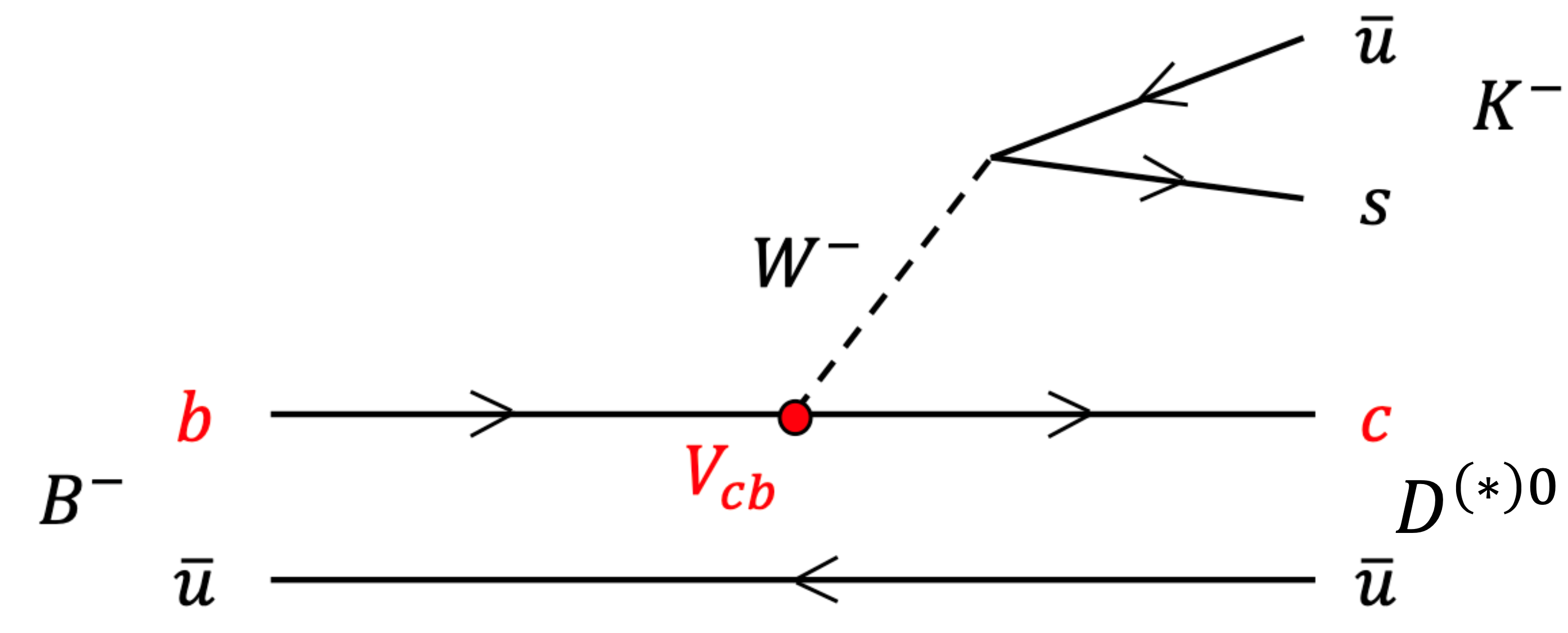}
    \includegraphics[width=0.33\textwidth]{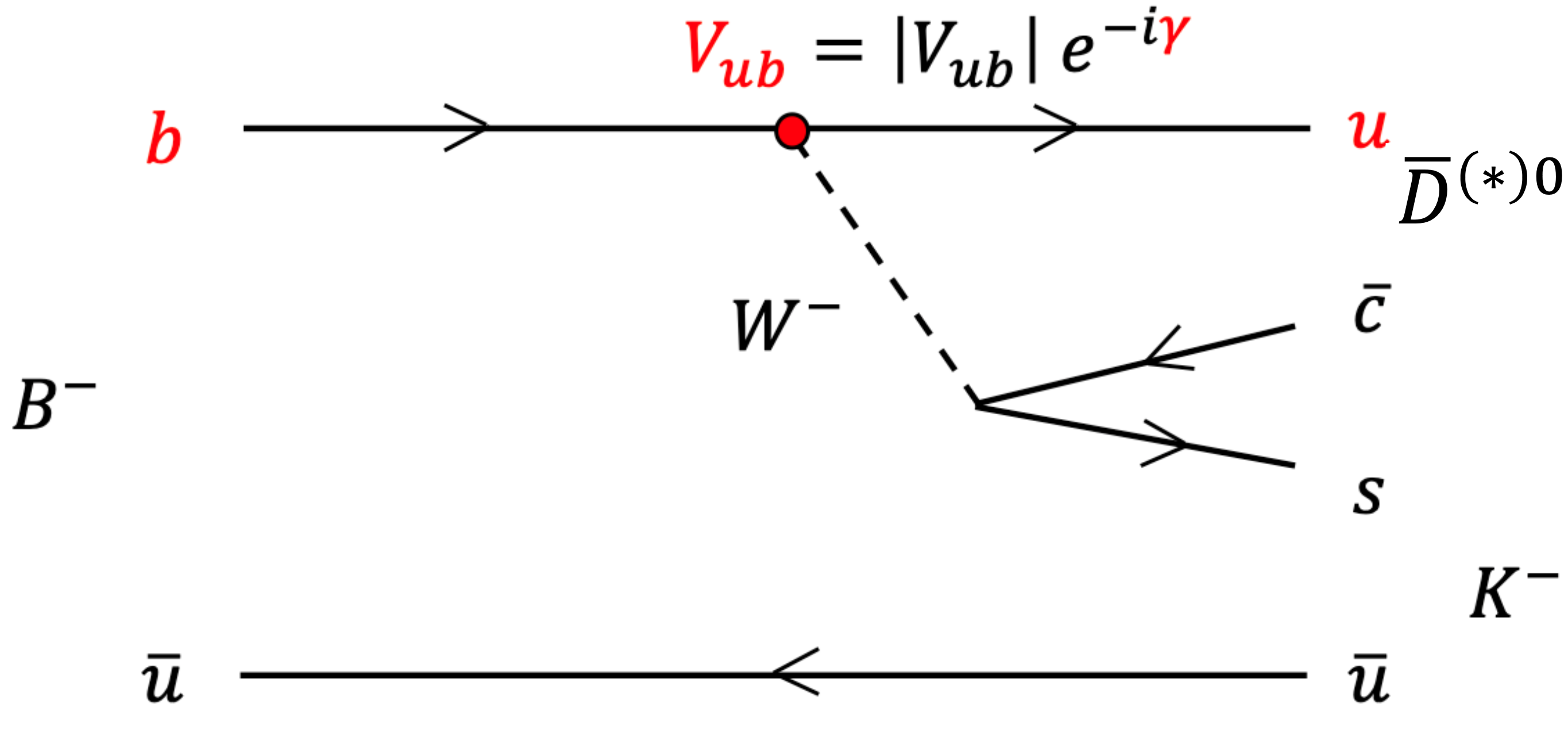}
    \caption{
    Description of the Feynman diagrams for the processes (left) $B^-\rightarrow D^{(*)0}K^-$ and (right) $B^-\rightarrow \bar{D}^{(*)0}K^-$.}
    \label{fig:triangle}
\end{figure}

The most powerful method for determining the angle $\gamma$ in decays dominated by tree-level contributions utilizes the $B^{\pm}\rightarrow D^{(*)}K^{\pm}$ decays, where $D^{(*)}$ represents an admixture of the $D^{(*)0}$ and $\bar{D}^{(*)0}$ states. 
Fig.~\ref{fig:triangle} shows the interference between $b\rightarrow c$ and $b\rightarrow u$ can give sensitivity to the angle $\gamma$~\cite{Gronau:1991dp}. Several methods are available to measure the angle $\gamma$ using decays such as $B^{\pm} \rightarrow D^{(*)}K^{\pm}$. The GLW method considers decays of the $D$ meson to $\CP$ eigenstates. The ADS approach requires Cabibbo-favored and doubly Cabibbo-suppressed (DCS) $D$ decays. 
BPGGSZ utilizes $D$ decays to self-conjugate final states, measuring the $\CP$ asymmetries over the phase space.

\begin{figure}
    \centering
    \includegraphics[width=0.45\textwidth]{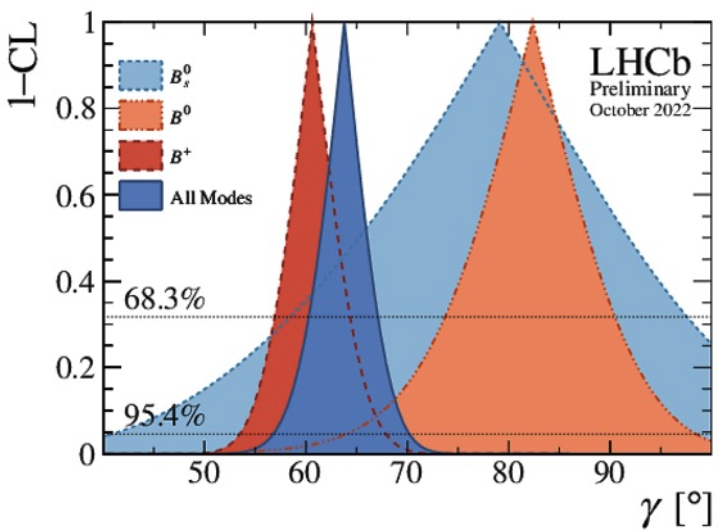}
    \includegraphics[width=0.45\textwidth]{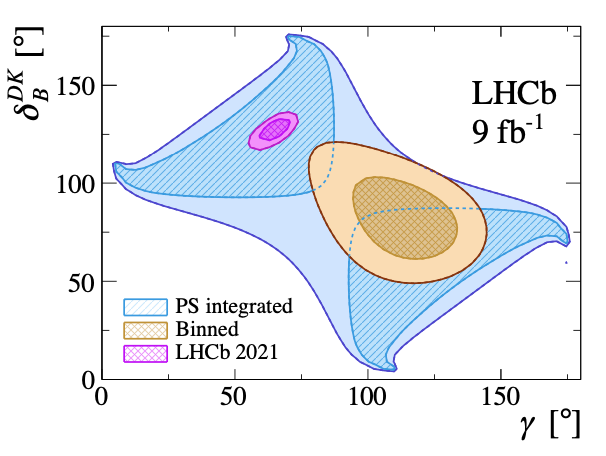}
    \caption{Combination (left) of the CKM angle $\gamma$ determined in $\lhcb$ and the confidence level (right) of $\gamma$ and $\delta_B^{DK}$ are shown, PS integrated results in the blue region and binned results in the brown region as well as \lhcb 2021 results in the pink region. }
    \label{fig:gammacombine}
\end{figure}

The first measurement of $\CP$ violation in the $B^{\pm} \rightarrow [K^+ K^- \pi^+ \pi^-]_D h^{\pm}$ mode, where $h$ is $K$ or $\pi$, is presented~\cite{LHCb:2023yjo}. 
It is the first measurements of the global $\CP$ asymmetries for this decay, and global measurements have been updated for the mode $B^{\pm} \rightarrow [\pi^+\pi^- \pi^+\pi^- ]_{D} h^{\pm}$. The $\CP$ observables obtained from the phase space (PS) integrated measurements can be interpreted in terms of the angle $\gamma$ and strong phase differences. The results are compatible with measurements using other decay channels and supersede the previous $B^{\pm} \rightarrow [\pi^+\pi^- \pi^+\pi^- ]_{D} h^{\pm}$ measurement~\cite{LHCb:2020hdx}, as shown on the right in Fig.~\ref{fig:gammacombine} with blue region. 

The analysis is also performed in bins of PS, which are optimized for sensitivity to local $\CP$ asymmetries. The analysis requires external information on charm-decay parameters~\cite{LHCb:2021dcr}, which are currently taken from an amplitude analysis of \lhcb data but can be updated in the future when direct measurements become available. That will allow the $CP$-violating observables to be determined in a model-independent fashion. A model-dependent value of $(116\pm12)^{\circ}$ is obtained. This result will be updated when the observables are re-evaluated using model-independent inputs. The precision is limited by the sample size and is expected to improve with future data from \lhcb, as shown on the right in Fig.~\ref{fig:gammacombine} with brown region.

The combination of measurements of the angle $\gamma$ is updated, including two new measurements related to $B$ decays published by the \lhcb in 2022 such as \mbox{$B^{\pm}\rightarrow [K^{\mp}\pi^{\pm}\pi^{\pm}\pi^{\mp}]_Dh^{\pm}$}~\cite{LHCb:2022nng}, \mbox{$B^{\pm}\rightarrow [h^{\pm}h'^{\mp}\pi^{0}]_Dh^{\pm}$}~\cite{LHCb:2021mmv} modes, as well as updates in the charm sector~\cite{LHCb:2022awq}. The determined value $(63.8^{+3.5}_{-3.7})^{\circ}$, as shown in Fig.~\ref{fig:gammacombine} (left), is compatible with the previous \lhcb combination~\cite{LHCb:2021dcr}, $(65.4^{+3.8}_{-4.2})^{\circ}$, and in excellent agreement with the global CKM fit predictions~\cite{Charles:2020dfl}. 
This is the most precise determination of the angle $\gamma$ from a single experiment~\cite{LHCb:2022awq}.

\section{\texorpdfstring{\CP}{CP} violation in charm sector}

The charm sector offers a unique probe to the up-type sector which is important on its own as well as complementary to searches in the strange and beauty sector. 
The asymmetries from the CKM matrix elements responsible for $\CP$ violation in charm decays are typically of the order of $10^{-4}-10^{-3}$ in the SM~\cite{GOLDEN1989501,Buccella:1994nf,Bianco:2003vb,Grossman:2006jg,Artuso:2008vf}. In the charm quark sector, the recent observation of $\CP$ violation has stimulated a wide discussion. 
The discovery used the difference of \CP asymmetries in $D\rightarrow K^- K^+$ and $D\rightarrow \pi^- \pi^+$ decays~\cite{LHCb:2019hro}. 
Further precise measurements are necessary and may resolve the theoretical debate on whether the observed value is consistent with the SM. 
For DCS decays, $\CP$ violation is highly suppressed within the SM~\cite{Grossman:2006jg,Bianco:2020hzf,Bergmann:1999pm}, thus its observation would indicate a manifestation of physics beyond the SM. Direct $\CP$ violation occurs when a given final state is produced through amplitudes with different weak phases, while the presence of different strong phases is also required. Final states are mainly reached via resonances in multi-body charm decays, providing an important source of strong-phase differences that vary across the two-dimensional Dalitz plot. This feature can enhance the sensitivity to $\CP$ asymmetries. 
The model-independent method is used for two multi-body $D$ decay measurements. Comparing the Dalitz distributions of $D^0$ and $\bar{D}^0$ meson decays provides a sensitive search for $\CP$ violation within the phase space of these decays. 

For the $D^0\rightarrow \pi^-\pi^+\pi^0$ decay, the unbinned model-independent method is used~\cite{LHCb:2023mwc}. The test quantifies localized sample differences, which can be converted into a p-value by comparing the nominal result to the distribution expected under the null hypothesis. The p-value is given, providing no indication of any $\CP$ violation in localized regions of the phase space. This measurement confirms that potential nuisance asymmetries can be neglected for data corresponding to Run I and II statistics.

A search for direct $\CP$ violation in the Cabibbo-suppressed decay $D_s^+ \rightarrow K^-K^+K^+$ and in the doubly Cabibbo-suppressed decay $D^+ \rightarrow K^-K^+K^+$ is reported~\cite{LHCb:2023qne}. The binned model-independent method is used. A variation of the origin $Miranda$ technique~\cite{BaBar:2008xzl,Bediaga:2009tr} is applied, dividing the PS in two-dimensional bins and computing the significance of the difference in the number of $D_{(s)}^+$ and $D_{(s)}^-$ candidates. A two-sample $\chi^2$ test is performed on the $D_{(s)}^+$ and $D_{(s)}^-$ samples. The resulting p-value from this test is defined as the probability of obtaining a test variable that is at least as high as the value observed, under the assumption of $\CP$ conservation. 
The results are given as p-values with respect to the null hypothesis of \CP conservation. They are found to be 13.3\% for the $D_s^+ \rightarrow K^- K^+ K^+$ channel and 31.6\% for the $D^+ \rightarrow K^-K^+K^+$ channel. The results are consistent with the hypothesis of no localized \CP violation in either channel. 
No evidence of \CP violation is found. This is the first search for $\CP$ violation in the Cabibbo-suppressed channel $D_s^+ \rightarrow K^- K^+ K^+$ and in the doubly Cabibbo-suppressed channel $D^+ \rightarrow K^- K^+ K^+$.

\section{summary}
The measurements of $\phi_s$ by \atlas, \cms and \lhcb are in agreement with the SM. Results based on decays with electrons in the final state are an important crosscheck.

The $\phi_s^{s\bar{s}s}$ measurement is consistent with and supersedes the previous measurement, and agrees with the SM expectation of a tiny $\CP$ violation. 

The direct measurement of the angle $\gamma$ in $B$ decays improves precision, with an uncertainty smaller than $4^{\circ}$. The precision will be further improved with other decay modes and more knowledge of charm hadronic parameters. 
 
The first search for $\CP$ violation in the Cabibbo-suppressed channel $D_s^+ \rightarrow K^-K^+K^+$ and in the doubly Cabibbo-suppressed channel $D^+ \rightarrow K^-K^+K^+$ has been presented. The results are consistent with the hypothesis of no localized \CP violation in either channel. There is also a new search of $\CP$ violation in the charm sector such as $D\rightarrow \pi\pi\pi^0$ channel with results being consistent with no $\CP$ violation hypothesis. 

\addcontentsline{toc}{section}{References}

\bibliographystyle{JHEP}
\bibliography{skeleton}

\end{document}

%% file: extradef.tex
\usepackage{xspace} 
\usepackage{upgreek}

\def\lhcb   {\mbox{LHCb}\xspace}
\def\atlas  {\mbox{ATLAS}\xspace}
\def\cms    {\mbox{CMS}\xspace}

\def\cdf    {\mbox{CDF}\xspace}
\def\dzero  {\mbox{D0}\xspace}

\def\CP                {{\ensuremath{C\!P}}\xspace}

\newcommand{\tev}{\aunit{Te\kern -0.1em V}\xspace}

%% file: skeleton.bbl
\providecommand{\href}[2]{#2}\begingroup\raggedright\begin{thebibliography}{10}

\bibitem{PhysRevLett.10.531}
N.~Cabibbo, \emph{Unitary symmetry and leptonic decays}, \href{https://doi.org/10.1103/PhysRevLett.10.531}{\emph{Phys. Rev. Lett.} {\bfseries 10} (1963) 531}.

\bibitem{Rosner:1996yb}
J.L.~Rosner, \emph{{CKM matrix and standard model CP violation}}, \href{https://doi.org/10.1016/S0920-5632(97)00423-4}{\emph{Nucl. Phys. B Proc. Suppl.} {\bfseries 59} (1997) 1} [\href{https://arxiv.org/abs/hep-ph/9612327}{{\ttfamily hep-ph/9612327}}].

\bibitem{ParticleDataGroup:2018ovx}
{\scshape Particle Data Group} collaboration, \emph{{Review of Particle Physics}}, \href{https://doi.org/10.1103/PhysRevD.98.030001}{\emph{Phys. Rev. D} {\bfseries 98} (2018) 030001}.

\bibitem{Charles:2020dfl}
J.~Charles, S.~Descotes-Genon, Z.~Ligeti, S.~Monteil, M.~Papucci, K.~Trabelsi et~al., \emph{{New physics in $B$ meson mixing: future sensitivity and limitations}}, \href{https://doi.org/10.1103/PhysRevD.102.056023}{\emph{Phys. Rev. D} {\bfseries 102} (2020) 056023} [\href{https://arxiv.org/abs/2006.04824}{{\ttfamily 2006.04824}}].

\bibitem{UTfit:2022hsi}
{\scshape UTfit} collaboration, \emph{{New UTfit Analysis of the Unitarity Triangle in the Cabibbo-Kobayashi-Maskawa scheme}}, \href{https://doi.org/10.1007/s12210-023-01137-5}{\emph{Rend. Lincei Sci. Fis. Nat.} {\bfseries 34} (2023) 37} [\href{https://arxiv.org/abs/2212.03894}{{\ttfamily 2212.03894}}].

\bibitem{Artuso:2015swg}
M.~Artuso, G.~Borissov and A.~Lenz, \emph{{CP violation in the $B_s^0$ system}}, \href{https://doi.org/10.1103/RevModPhys.88.045002}{\emph{Rev. Mod. Phys.} {\bfseries 88} (2016) 045002} [\href{https://arxiv.org/abs/1511.09466}{{\ttfamily 1511.09466}}].

\bibitem{ATLAS:2020lbz}
{\scshape ATLAS} collaboration, \emph{{Measurement of the $CP$-violating phase $\phi_s$ in $B^0_s \to J/\psi\phi$ decays in ATLAS at 13 TeV}}, \href{https://doi.org/10.1140/epjc/s10052-021-09011-0}{\emph{Eur. Phys. J. C} {\bfseries 81} (2021) 342} [\href{https://arxiv.org/abs/2001.07115}{{\ttfamily 2001.07115}}].

\bibitem{CMS:2020efq}
{\scshape CMS} collaboration, \emph{{Measurement of the $CP$-violating phase $\phi_\mathrm{s}$ in the B$^0_\mathrm{s}\to$ J$/\psi\, \phi$(1020) $\to \mu^+\mu^-$K$^+$K$^-$ channel in proton-proton collisions at $\sqrt{s} =$ 13 TeV}}, \href{https://doi.org/10.1016/j.physletb.2021.136188}{\emph{Phys. Lett. B} {\bfseries 816} (2021) 136188} [\href{https://arxiv.org/abs/2007.02434}{{\ttfamily 2007.02434}}].

\bibitem{LHCb:2021wte}
{LHCb collaboration}, \emph{{First measurement of the $C\!P$-violating phase in ${{B} ^0_{s}} \!\rightarrow {{J /\psi }} (\rightarrow e ^+e ^-$)$\phi $ decays}}, \href{https://doi.org/10.1140/epjc/s10052-021-09711-7}{\emph{Eur. Phys. J. C} {\bfseries 81} (2021) 1026} [\href{https://arxiv.org/abs/2105.14738}{{\ttfamily 2105.14738}}].

\bibitem{LHCb:2019nin}
{LHCb collaboration}, \emph{{Updated measurement of time-dependent CP-violating observables in $B^{0}_{s}\to J/\psi K^+ K^-$ decays}}, \href{https://doi.org/10.1140/epjc/s10052-019-7159-8}{\emph{Eur. Phys. J. C} {\bfseries 79} (2019) 706} [\href{https://arxiv.org/abs/1906.08356}{{\ttfamily 1906.08356}}].

\bibitem{HFLAV:2022esi}
{\scshape HFLAV} collaboration, \emph{{Averages of b-hadron, c-hadron, and \ensuremath{\tau}-lepton properties as of 2021}}, \href{https://doi.org/10.1103/PhysRevD.107.052008}{\emph{Phys. Rev. D} {\bfseries 107} (2023) 052008} [\href{https://arxiv.org/abs/2206.07501}{{\ttfamily 2206.07501}}].

\bibitem{LHCb:2023exl}
{LHCb collaboration}, \emph{{Precision Measurement of CP Violation in the Penguin-Mediated Decay Bs0\textrightarrow{}\ensuremath{\phi}\ensuremath{\phi}}}, \href{https://doi.org/10.1103/PhysRevLett.131.171802}{\emph{Phys. Rev. Lett.} {\bfseries 131} (2023) 171802} [\href{https://arxiv.org/abs/2304.06198}{{\ttfamily 2304.06198}}].

\bibitem{LHCb:2013xyz}
{LHCb collaboration}, \emph{{First measurement of the CP-violating phase in $B_s^0 \to \phi \phi$ decays}}, \href{https://doi.org/10.1103/PhysRevLett.110.241802}{\emph{Phys. Rev. Lett.} {\bfseries 110} (2013) 241802} [\href{https://arxiv.org/abs/1303.7125}{{\ttfamily 1303.7125}}].

\bibitem{LHCb:2014vmq}
{LHCb collaboration}, \emph{{Measurement of CP violation in $B_s^0 \to \phi \phi$ decays}}, \href{https://doi.org/10.1103/PhysRevD.90.052011}{\emph{Phys. Rev. D} {\bfseries 90} (2014) 052011} [\href{https://arxiv.org/abs/1407.2222}{{\ttfamily 1407.2222}}].

\bibitem{LHCb:2019jgw}
{LHCb collaboration}, \emph{{Measurement of CP violation in the $ {B}_s^0\to \phi \phi $ decay and search for the $B^0\rightarrow \phi\phi$ decay}}, \href{https://doi.org/10.1007/JHEP12(2019)155}{\emph{JHEP} {\bfseries 12} (2019) 155} [\href{https://arxiv.org/abs/1907.10003}{{\ttfamily 1907.10003}}].

\bibitem{Cheng:2009mu}
H.-Y.~Cheng and C.-K.~Chua, \emph{{{QCD} Factorization for Charmless Hadronic $B_s$ Decays Revisited}}, \href{https://doi.org/10.1103/PhysRevD.80.114026}{\emph{Phys. Rev. D} {\bfseries 80} (2009) 114026} [\href{https://arxiv.org/abs/0910.5237}{{\ttfamily 0910.5237}}].

\bibitem{Beneke:2006hg}
M.~Beneke, J.~Rohrer and D.~Yang, \emph{{Branching fractions, polarisation and asymmetries of B ---\ensuremath{>} VV decays}}, \href{https://doi.org/10.1016/j.nuclphysb.2007.03.020}{\emph{Nucl. Phys. B} {\bfseries 774} (2007) 64} [\href{https://arxiv.org/abs/hep-ph/0612290}{{\ttfamily hep-ph/0612290}}].

\bibitem{Raidal:2002ph}
M.~Raidal, \emph{{CP asymmetry in B ---\ensuremath{>} phi K(S) decays in left-right models and its implications on B(s) decays}}, \href{https://doi.org/10.1103/PhysRevLett.89.231803}{\emph{Phys. Rev. Lett.} {\bfseries 89} (2002) 231803} [\href{https://arxiv.org/abs/hep-ph/0208091}{{\ttfamily hep-ph/0208091}}].

\bibitem{Brod:2013sga}
J.~Brod and J.~Zupan, \emph{{The ultimate theoretical error on $\gamma$ from $B \to DK$ decays}}, \href{https://doi.org/10.1007/JHEP01(2014)051}{\emph{JHEP} {\bfseries 01} (2014) 051} [\href{https://arxiv.org/abs/1308.5663}{{\ttfamily 1308.5663}}].

\bibitem{Gronau:1991dp}
M.~Gronau and D.~Wyler, \emph{{On determining a weak phase from CP asymmetries in charged B decays}}, \href{https://doi.org/10.1016/0370-2693(91)90034-N}{\emph{Phys. Lett. B} {\bfseries 265} (1991) 172}.

\bibitem{LHCb:2022nng}
{LHCb collaboration}, \emph{{Measurement of the CKM angle $\gamma$ with $ B^\pm \to D[K^\mp \pi^\pm \pi^\pm \pi^\mp] h^\pm$ decays using a binned phase-space approach}}, \href{https://doi.org/10.1007/JHEP07(2023)138}{\emph{JHEP} {\bfseries 07} (2023) 138} [\href{https://arxiv.org/abs/2209.03692}{{\ttfamily 2209.03692}}].

\bibitem{LHCb:2021mmv}
{LHCb collaboration}, \emph{{Constraints on the CKM angle $\gamma$ from $B^\pm\rightarrow Dh^\pm$ decays using $D\rightarrow h^\pm h^{\prime\mp}\pi^0$ final states}}, \href{https://doi.org/10.1007/JHEP07(2022)099}{\emph{JHEP} {\bfseries 07} (2022) 099} [\href{https://arxiv.org/abs/2112.10617}{{\ttfamily 2112.10617}}].

\bibitem{LHCb:2022awq}
{LHCb collaboration}, \emph{{Simultaneous determination of the CKM angle $\gamma$ and parameters related to mixing and $CP$ violation in the charm sector}} .

\bibitem{LHCb:2021dcr}
{LHCb collaboration}, \emph{{Simultaneous determination of CKM angle $\gamma$ and charm mixing parameters}}, \href{https://doi.org/10.1007/JHEP12(2021)141}{\emph{JHEP} {\bfseries 12} (2021) 141} [\href{https://arxiv.org/abs/2110.02350}{{\ttfamily 2110.02350}}].

\bibitem{LHCb:2023yjo}
{LHCb collaboration}, \emph{{A study of $C\!P$ violation in the decays $B^\pm\to[K^+K^-\pi^+\pi^-]_D h^{\pm}$ ($h = K, \pi$) and $B^\pm\to[\pi^+\pi^-\pi^+\pi^-]_D h^{\pm}$}}, \href{https://doi.org/10.1140/epjc/s10052-023-11560-5}{\emph{Eur. Phys. J. C} {\bfseries 83} (2023) 547} [\href{https://arxiv.org/abs/2301.10328}{{\ttfamily 2301.10328}}].

\bibitem{LHCb:2020hdx}
{LHCb collaboration}, \emph{{Measurement of CP observables in $B^\pm \to D^{(*)} K^\pm$ and $B^\pm \to D^{(*)} \pi^\pm$ decays using two-body $D$ final states}}, \href{https://doi.org/10.1007/JHEP04(2021)081}{\emph{JHEP} {\bfseries 04} (2021) 081} [\href{https://arxiv.org/abs/2012.09903}{{\ttfamily 2012.09903}}].

\bibitem{GOLDEN1989501}
M.~Golden and B.~Grinstein, \emph{Enhanced cp violations in hadronic charm decays}, \href{https://doi.org/https://doi.org/10.1016/0370-2693(89)90353-5}{\emph{Physics Letters B} {\bfseries 222} (1989) 501}.

\bibitem{Buccella:1994nf}
F.~Buccella, M.~Lusignoli, G.~Miele, A.~Pugliese and P.~Santorelli, \emph{{Nonleptonic weak decays of charmed mesons}}, \href{https://doi.org/10.1103/PhysRevD.51.3478}{\emph{Phys. Rev. D} {\bfseries 51} (1995) 3478} [\href{https://arxiv.org/abs/hep-ph/9411286}{{\ttfamily hep-ph/9411286}}].

\bibitem{Bianco:2003vb}
S.~Bianco, F.L.~Fabbri, D.~Benson and I.~Bigi, \emph{{A Cicerone for the physics of charm}}, \href{https://doi.org/10.1393/ncr/i2003-10003-1}{\emph{Riv. Nuovo Cim.} {\bfseries 26} (2003) 1} [\href{https://arxiv.org/abs/hep-ex/0309021}{{\ttfamily hep-ex/0309021}}].

\bibitem{Grossman:2006jg}
Y.~Grossman, A.L.~Kagan and Y.~Nir, \emph{{New physics and CP violation in singly Cabibbo suppressed D decays}}, \href{https://doi.org/10.1103/PhysRevD.75.036008}{\emph{Phys. Rev. D} {\bfseries 75} (2007) 036008} [\href{https://arxiv.org/abs/hep-ph/0609178}{{\ttfamily hep-ph/0609178}}].

\bibitem{Artuso:2008vf}
M.~Artuso, B.~Meadows and A.A.~Petrov, \emph{{Charm Meson Decays}}, \href{https://doi.org/10.1146/annurev.nucl.58.110707.171131}{\emph{Ann. Rev. Nucl. Part. Sci.} {\bfseries 58} (2008) 249} [\href{https://arxiv.org/abs/0802.2934}{{\ttfamily 0802.2934}}].

\bibitem{LHCb:2019hro}
{LHCb collaboration}, \emph{{Observation of CP Violation in Charm Decays}}, \href{https://doi.org/10.1103/PhysRevLett.122.211803}{\emph{Phys. Rev. Lett.} {\bfseries 122} (2019) 211803} [\href{https://arxiv.org/abs/1903.08726}{{\ttfamily 1903.08726}}].

\bibitem{Bianco:2020hzf}
S.~Bianco and I.I.~Bigi, \emph{{2019/20 lessons from $\tau (\Omega_c^0)$ and $\tau (\Xi_c^0)$ and CP asymmetry in charm decays}}, \href{https://doi.org/10.1142/S0217751X20300136}{\emph{Int. J. Mod. Phys. A} {\bfseries 35} (2020) 2030013} [\href{https://arxiv.org/abs/2001.06908}{{\ttfamily 2001.06908}}].

\bibitem{Bergmann:1999pm}
S.~Bergmann and Y.~Nir, \emph{{New physics effects in doubly Cabibbo suppressed D decays}}, \href{https://doi.org/10.1088/1126-6708/1999/09/031}{\emph{JHEP} {\bfseries 09} (1999) 031} [\href{https://arxiv.org/abs/hep-ph/9909391}{{\ttfamily hep-ph/9909391}}].

\bibitem{LHCb:2023mwc}
{LHCb collaboration}, \emph{{Search for CP violation in the phase space of D$^{0}$\textrightarrow{} \ensuremath{\pi}$^{-}$\ensuremath{\pi}$^{+}$\ensuremath{\pi}$^{0}$ decays with the energy test}}, \href{https://doi.org/10.1007/JHEP09(2023)129}{\emph{JHEP} {\bfseries 09} (2023) 129} [\href{https://arxiv.org/abs/2306.12746}{{\ttfamily 2306.12746}}].

\bibitem{LHCb:2023qne}
{LHCb collaboration}, \emph{{Search for $\it{CP}$ violation in $D_{(s)}^{+}\rightarrow K^{-}K^{+}K^{+}$ decays}}, \href{https://doi.org/10.1007/JHEP07(2023)067}{\emph{JHEP} {\bfseries 07} (2023) 067} [\href{https://arxiv.org/abs/2303.04062}{{\ttfamily 2303.04062}}].

\bibitem{BaBar:2008xzl}
{\scshape BaBar} collaboration, \emph{{Search for CP Violation in Neutral D Meson Cabibbo-suppressed Three-body Decays}}, \href{https://doi.org/10.1103/PhysRevD.78.051102}{\emph{Phys. Rev. D} {\bfseries 78} (2008) 051102} [\href{https://arxiv.org/abs/0802.4035}{{\ttfamily 0802.4035}}].

\bibitem{Bediaga:2009tr}
I.~Bediaga, I.I.~Bigi, A.~Gomes, G.~Guerrer, J.~Miranda and A.C.d.~Reis, \emph{{On a CP anisotropy measurement in the Dalitz plot}}, \href{https://doi.org/10.1103/PhysRevD.80.096006}{\emph{Phys. Rev. D} {\bfseries 80} (2009) 096006} [\href{https://arxiv.org/abs/0905.4233}{{\ttfamily 0905.4233}}].

\end{thebibliography}\endgroup
